\documentclass[twoside,a4paper,11pt]{torus2012}
\usepackage{graphicx}
\usepackage{hyperref}
\usepackage{movie15}
\begin{document}
\pagenumbering{arabic}
\pagestyle{myheadings}
\thispagestyle{empty}
{\flushleft\includegraphics[width=\textwidth,bb=58 650 590 680]{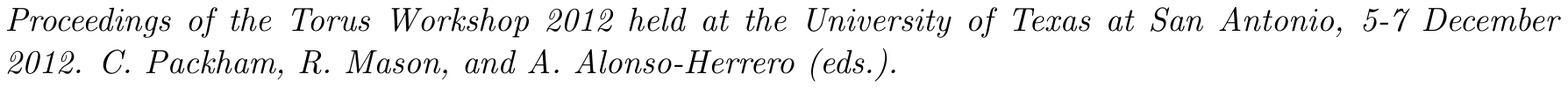}}
\vspace*{0.2cm}
\begin{flushleft}
{\bf {\LARGE
%
Looking for extended nuclear emission in a sample of nearby QSOs
%
}\\
\vspace*{1cm}
%
M. Mart\'{\i}nez-Paredes$^{1}$,
I. Aretxaga$^{1}$, 
A. Alonso-Herrero$^{2}$,
C. Packham$^{3}$
and CanariCam Piratas team
%
}\\
\vspace*{0.5cm}
%
$^{1}$
Instituto Nacional de Astrof\'isica, \'Optica y Electr\'onica, Tonantzintla, Puebla, Mexico\\
$^{2}$
Instituto de F\'isica de Cantabria, CSIC-UC Santander, Spain \\
$^{3}$
The University of Texas at San Antonio, San Antonio, USA
%
\end{flushleft}
%
\markboth{
Extended nuclear emission in a sample of nearby QSO
}{ 
%
Mart\'{\i}nez-Paredes et al.
%
}
\thispagestyle{empty}
\vspace*{0.4cm}
\begin{minipage}[l]{0.09\textwidth}
\ 
\end{minipage}
\begin{minipage}[r]{0.9\textwidth}
\vspace{1cm}
\section*{Abstract}{\small
%
We present MIR sub-arcsec observations of Mrk 1383, a nearby QSO, part of a sample of nearby QSOs 
currently being observed with the 10.4 m Gran Telescopio de Canarias (GTC). The radial profile of 
the QSO is analyzed and we find that most of the MIR emission is unresolved
($\theta\sim 0.4"$, 600 pc) and consistent 
with previous measurements from Spitzer ($\theta \sim 4"$). We derive the spectral energy distribution (SED) 
of the nuclear emission, relatively uncontaminated by starlight, combining measurements from the 
literature as a first step towards characterizing the torus parameters in this QSO.
%
\normalsize}
\end{minipage}
%
%
%
\section{Introduction \label{intro}}

The unified model for active galactic nuclei (AGN) proposes the ubiquitous presence of 
obscuring tori around their nuclei, with type 1 and type  2 AGN being
intrinsically similar (\cite{Antonucci}). 
The infrared (IR) range (and particularly the mid-IR)  is key to characterize the torus, since dust reprocesses 
the optical and ultraviolet radiation generated in the accretion process and re-emits it in this range. However, 
considering the small torus size ($<10$ pc in the case of Seyfert galaxies), high angular resolution turns to be 
essential to isolate as much as possible its emission.
During the last two decades much progress has been made towards understanding the properties of the molecular 
dusty torus. For example, some authors assumed a uniform dust density distribution to simplify the modeling (\cite{Pier1}, \cite{Pier2}). 
To solve the various discrepancies between observations and models, an intensive search for an alternative torus 
geometry has been carried out in the last years. The clumpy torus models propose that the dust is distributed 
in clumps, instead of homogeneously filling the torus volume (\cite{Nenkova},\cite{Honig}, \cite{Schartman}). These models are making significant progress in 
accounting for the MIR emission of AGN (\cite{Mason}, \cite{Horst}, \cite{Nikutta}, \cite{Ramos}, \cite{Honig2} and \cite{Alonso}). Little is known about the 
extent of the MIR emission in QSOs, 
the distribution of the polycyclic aromatic hydrocarbons (PAH) features detected in recent Spitzer spectroscopy (\cite{Veilleux}), 
and whether the unresolved emission is entirely due to dust-reprocessing by dusty tori as those detected in other 
types of AGN. Thus, it is important to characterize the emission of the brightest nearby QSOs in the MIR. 

\section{Sample definition and observations}

\begin{figure}
\center
\includegraphics[scale=0.2]{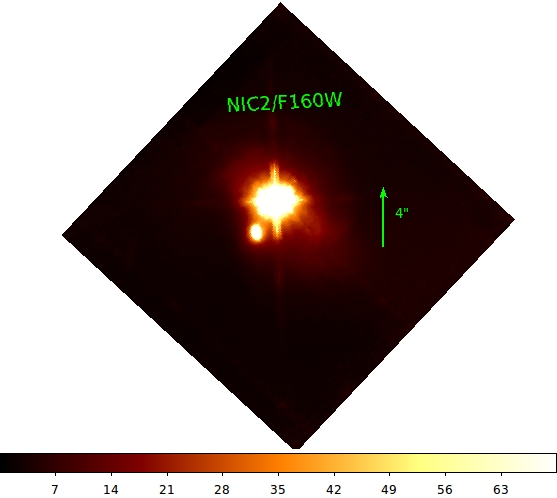} ~
\includegraphics[scale=0.2]{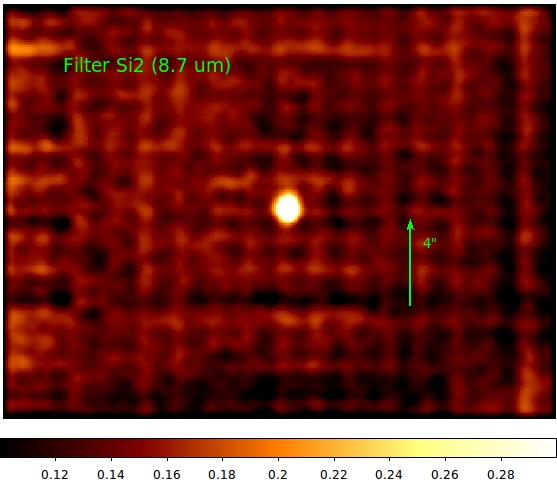} ~
\includegraphics[scale=0.2]{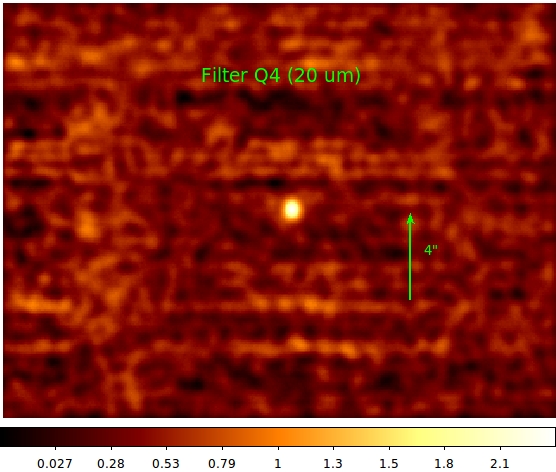} 
\caption{\label{fig1}  Left to right, image of Mrk 1383 at 1.6 $\mu$m (from HST archive), 8.7 $\mu$m and 20 $\mu$m (our CanariCam observations).
}
\end{figure}

Mrk 1383 ($z=0.079$) is part of a sample of 15 nearby QSOs being observed with CanariCam on the 10.4m 
Gran Telescopio de Canarias (GTC) in Spain, to study the prevalence of dusty tori and the significance and 
location of starbursts (SBs)  around powerful AGN. The sample was selected by redshift ($z<0.1$) and IR brightness 
$N>50$ mJy in order to secure detections at high angular-resolution
($\theta\sim 0.3-0.4"$). Additionally, we require that objects 
in our sample have the highest X-ray luminosities among AGN
($L$($2-10$ keV) $>10^{43}$ erg/s).
Mrk 1383 was imaged in March 2012 with the 10.4 m GTC at 8.7 and 20 $\mu$m, with a precipitable water vapor $PWV=3.5-3.9$ mm. 
The data reduction was performed with the  CanariCam pipeline of
\cite{Gonzalez}, and the images were flux calibrated with infrared standard stars 
from the Cohen's catalogue (\cite{Cohen}). Details of the observations can be found in Table~\ref{tab1} and Fig.~\ref{fig1}.

\begin{table}[ht] 
\caption{Details of the observations and measured fluxes within $0.72"$ apertures.} 
\center
\begin{minipage}{1.\textwidth}
\center
\begin{tabular}{ccccc} 
\hline\hline 
Filter & $\lambda$ ($\mu$m) & Time$_{exposure}$(sec) & FWHM
(arcsec)&$F_{\lambda}$ (mJy)\\ [0.5ex]   
\hline 
Si2 & 8.7 & 220 & 0.4 &$62.5\pm5.0$\\ 
Q4 & 20 & 220 & 0.48 &$232.2\pm49.5$ \\ [1ex]  
\hline
\end{tabular} 
\end{minipage}
\label{tab1} 
\end{table}

\section{Analysis}
\begin{figure}
\center
\includegraphics[scale=1]{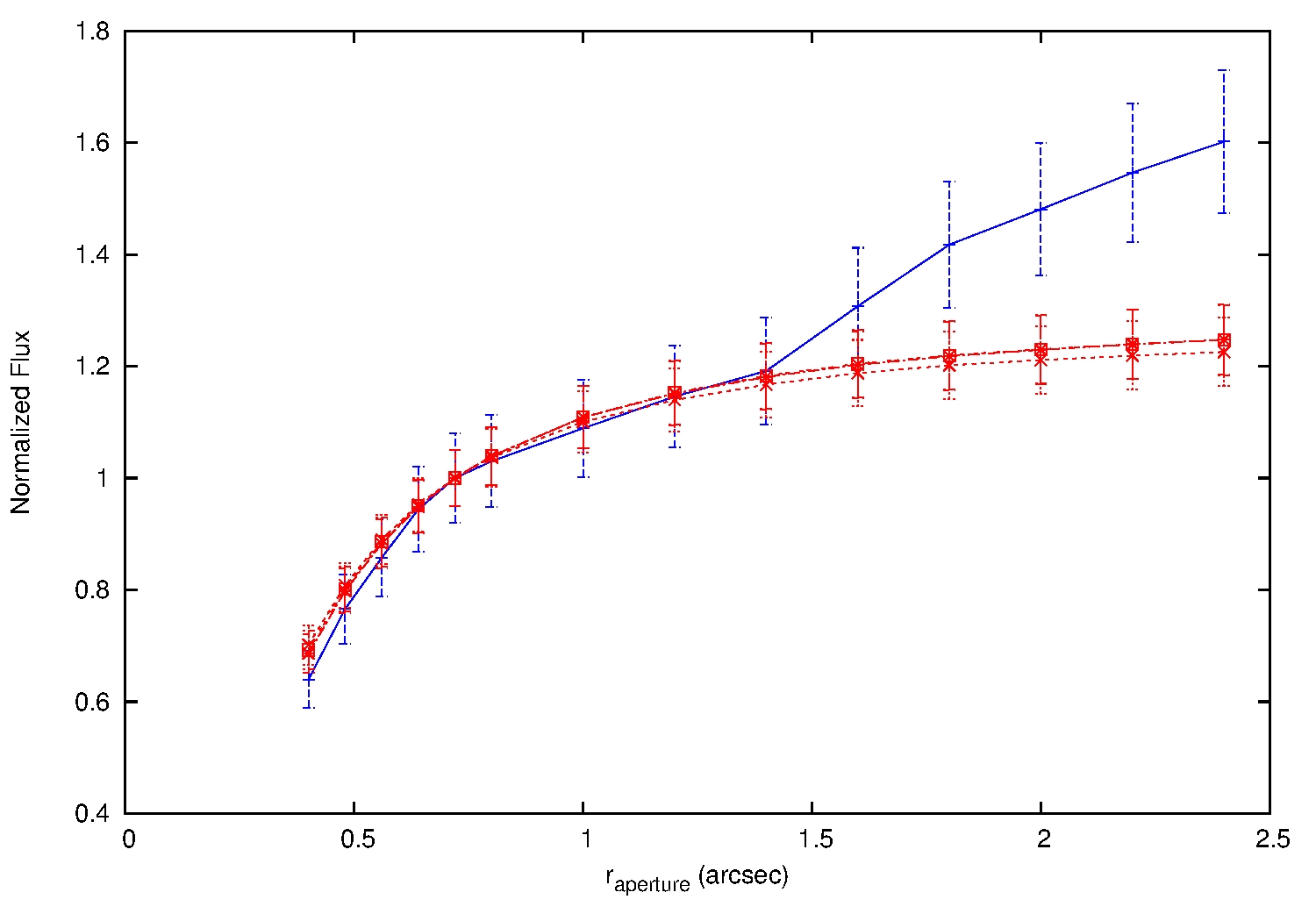} 
\caption{\label{fig2}  GTC/CanariCam radial profile at $8.7\,\mu$m of
  Mrk 1383 (blue 
  symbols and   line) compared to a
  standard star (3 repeats, red symbols and lines), acquired immediately after the galaxy observations.
}
\end{figure}

\begin{figure}
\center
\includegraphics[scale=0.95]{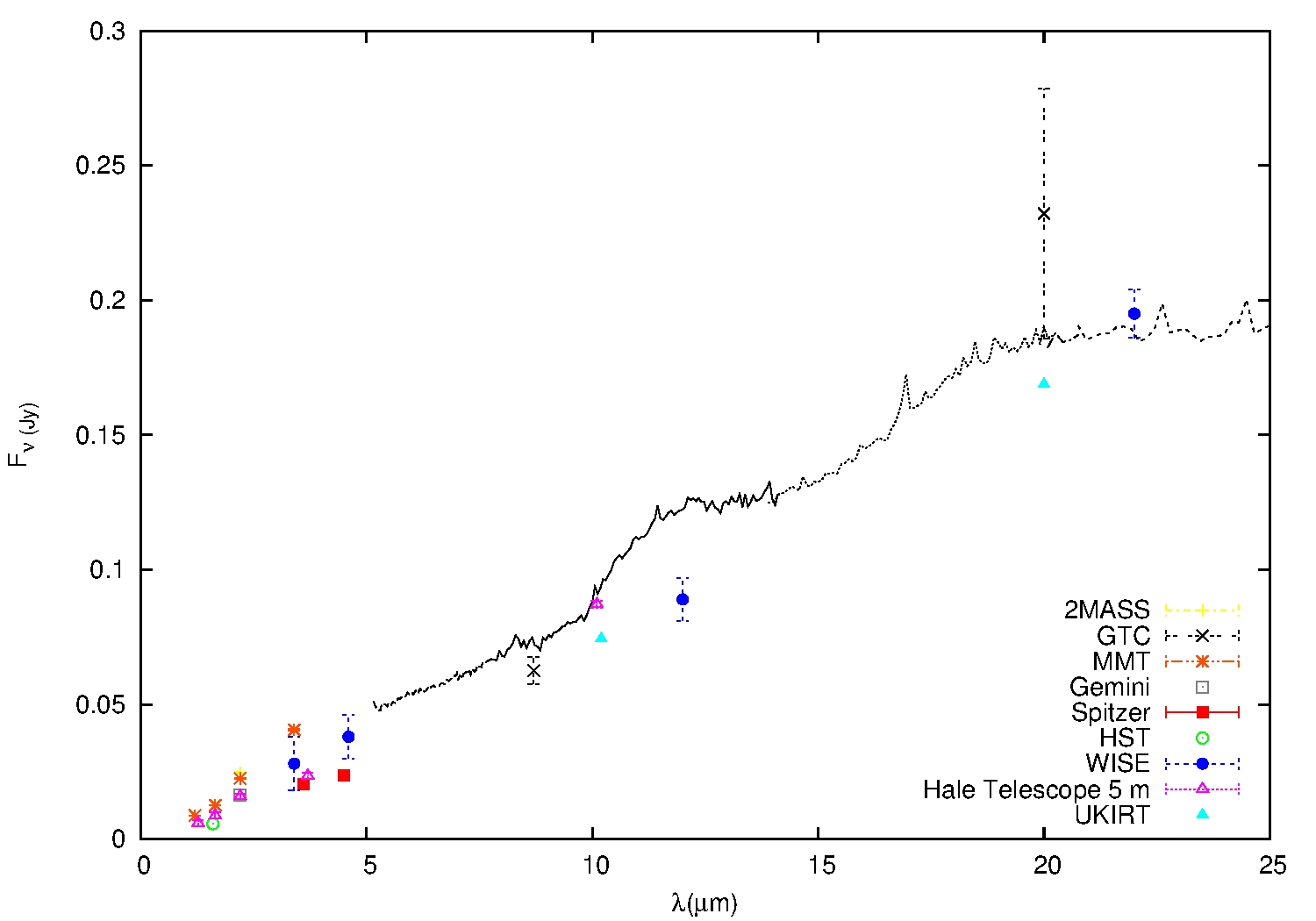} 
\caption{\label{fig3} Observed SED of Mrk 1383. The data from 2MASS, MMT and UKIRT and Hale Telescope are taken from references \cite{Elvis} and \cite{Neugebauer}. 
The IRAC and CanariCam photometry is from this work. The red line is
the Spitzer/IRS spectrum.}
\end{figure}

With the aim of looking for extended emission in Mrk 1383 we did aperture photometry on 
the target image using the phot/IRAF task varying the aperture from $0.4$ to $2.4"$ and compared 
it to the aperture photometry of the standard star. In Fig.~\ref{fig2} we show the radial profiles of 
the star acquired in 3 consecutive exposures and compare these with the radial profile of the galaxy. 
The radial profiles are very similar and we conclude that Mrk 1383 does not have prominent extended emission within the inner $1.5"$ and, 
therefore, its emission is mostly unresolved.
To construct a well-sampled IR SED (Fig.~\ref{fig3}), we compile near-IR high spatial resolution nuclear fluxes 
from the literature. The flux at 20 $\mu$m is larger than the flux from the Spitzer spectrum and the photometry 
obtain with UKIRT at the same wavelength. However, the observation conditions at this wavelength were not ideal 
and new observations are scheduled.

\section{Discussion}
We report subarcsecond resolution mid-IR fluxes for Mrk 1383. These nuclear fluxes, in combination with published 
near-IR measurements are used to construct the nuclear SED. We are currently analyzing and fitting  this SED with the
clumpy dusty torus models of \cite{Nenkova} and \cite{Nenkova2}.  This will allow us to put tight constraints on torus model parameters such as the 
viewing angle $i$, the radial thickness of the torus $Y$, the angular
size of the cloud distribution $\sigma_{\rm torus}$, and the average
number of clouds along radial equatorial rays $N$.

%

%
\end{document}